\documentclass[conference]{IEEEtran}
\normalsize
\ifCLASSINFOpdf
\else
\fi

\usepackage{amsthm}
\usepackage{amsmath}
\usepackage{amssymb}
\usepackage{graphicx}
\usepackage{esint}
\usepackage{amsfonts}
\usepackage{cite}
\usepackage{balance}
\usepackage{caption}
\usepackage{subcaption}
\usepackage{epstopdf}
\usepackage{color}
\makeatletter
\usepackage{amssymb}
\usepackage{gensymb}
\usepackage{lipsum}% http://ctan.org/pkg/lipsum
\usepackage{algorithm}% http://ctan.org/pkg/algorithm
\usepackage{algpseudocode}% http://ctan.org/pkg/algorithmicx
\usepackage[compatibility=false]{caption}% http://ctan.org/pkg/caption
\usepackage[bottom]{footmisc}
\usepackage[left=0.99in, right=0.99in, top=0.75in, bottom=0.75in]{geometry}
\theoremstyle{plain}

\makeatother

\providecommand{\theoremname}{Theorem}

\theoremstyle{plain}

\makeatother

\providecommand{\lemmaname}{Lemma}
\begin{document}
\title{\hspace{0cm}A Low-Overhead Hierarchical Beam-tracking Algorithm for THz Wireless Systems}

\author{%

\authorblockN{Giorgos Stratidakis, Alexandros--Apostolos A. Boulogeorgos\authorrefmark{\star}, Angeliki~Alexiou\authorrefmark{\star}}

\authorblockA{\authorrefmark{\star}\footnotesize  Department of Digital Systems, University of Piraeus, Piraeus 18534,

Greece. (E-mails: al.boulogeorgos@ieee.org, alexiou@unipi.gr).}

}
\author{

Giorgos Stratidakis$^{\star}$, Georgia D. Ntouni$^{\dagger}$, Alexandros--Apostolos A. Boulogeorgos$^{\star}$, \\ Dimitrios Kritharidis$^{\dagger}$, and~Angeliki~Alexiou$^{\star}$
\\

\begin{normalsize} 

$^{\star}$Department of Digital Systems, University of Piraeus, Piraeus 18534, Greece.
\end{normalsize}

\\
\begin{normalsize} 
$^{\dagger}$Intracom Telecom, 19.7 km Markopoulou, Ave., 190 02 Peania, Greece.
\end{normalsize}
\\

\begin{normalsize} 

E-mails: \{giostrat, alexiou\}@unipi.gr, \{gntouni, dkri\}@intracom-telecom.com, al.boulogeorgos@ieee.org.

\end{normalsize}

}

\maketitle	

\begin{abstract}
In this paper, a novel hierarchical beam-tracking approach, which is suitable for terahertz (THz) wireless systems, is presented. The main idea is to employ a prediction based algorithm with a multi-resolution codebook, in order to decrease the required overhead of tracking and increase its robustness. The efficiency of the algorithm is evaluated in terms of the average number of pilots and mean square error (MSE) and is compared with the corresponding performance of the fast channel tracking (FCT) algorithm. Our results highlight the superiority of the proposed approach in comparison with FCT, in terms of tracking efficiency with low overhead.

\end{abstract}


\section{Introduction}\label{S:Intro}

The key enabling technology for wireless terahertz (THz) communications is beamforming (BF), which is employed at both the access point (AP) and the user equipment (UE)~\cite{A:THz_Communications_An_Array_of_subarrays_solutions,Boulogeorgos2018,Papasotiriou2020,Boulogeorgos2020,C:UserAssociationInUltraDenseTHzNetworks,Boulogeorgos2019a}. However, BF comes with the necessity of accurate beam pointing in order to guarantee perfect transceiver antenna alignment and  support UE mobility~\cite{WP:Wireless_Thz_system_architecture_for_networks_beyond_5G}. As a consequence, several beam-tracking approaches have been presented for directional communication systems operating in lower frequency bands. The problem with adopting the conventional beam-tracking approach is the requirement for an unaffordable pilot overhead ~\cite{C:Time-domain_synchronous_OFDM_based_on_simultaneous_multi_channel_reconstruction}.
As a result, novel smart beam-tracking schemes are needed to avoid antenna misalignment~\cite{AlexandrosApostolosA.Boulogeorgos2020,Kokkoniemi2020}. 

Three main beam-tracking approaches for millimeter wave and THz wireless systems have been reported in the open technical literature, namely codebook-based, perturbation and prediction methods (see for example~\cite{C:Beampattern-based_tracking_for_mmW_communication_systems,
A:Channel_estimation_and_hybrid_precoding_for_mmW_cellular_systems,
Zhang2016,Va2016,Gao2017} and references therein). In more detail, in~\cite{C:Beampattern-based_tracking_for_mmW_communication_systems}, an algorithm that estimates the angle of arrival (AoA) by using variations in the radiation pattern of the beam as a function of the AoA is presented, whereas, in~\cite{A:Channel_estimation_and_hybrid_precoding_for_mmW_cellular_systems}, a  hierarchical multi-resolution codebook is designed to construct training beamforming vectors with different beamwidths. 
Likewise, in~\cite{Zhang2016}, a Kalman-filter in combination with an abrupt change detection were utilized for pencil-beam tracking, while in~\cite{Va2016}, an extended Kalman filter was employed that requires only one measurement of the single beam-pair to track the propagation path. Moreover, in~\cite{Gao2017},  the authors reported a beam-tracking prediction approach, which guarantees high-accuracy with low latency. However, the core assumption of this approach is that the UE follows linear motion. Thus, in a different scenario, this approach fails to provide the required accuracy under a low overhead specification. Finally, in~\cite{Stratidakis2019}, the authors proposed a cooperation-aided prediction-based beam-tracking approach, which relaxes the linearity motion demand of~\cite{Gao2017}.

 The disadvantages of the above mentioned approaches are that the codebook-based ones are limited by the resolution of the codebook, whereas the perturbation approaches require lengthy beam training, which incurs large overhead as well as delay. Finally, the prediction approaches suffer from low accuracy and usually support a small set of UE movements.
 %\\
%\indent 
To overcome the aforementioned restrictions, in this paper, a novel hierarchical beam-tracking approach, suitable for THz wireless systems is presented. This approach combines hierarchical codebook search with a location prediction algorithm. Both the hierarchical codebook and the location prediction are used to track the UE with low pilot overhead. The prediction algorithm is based on the observation that all motions can be described as multiple smaller linear motions. Therefore, by increasing the frequency of the direction estimations (estimations/sec), the prediction becomes more accurate. As a result, the tracking algorithm can reduce its pilot overhead significantly.
Finally, a hierarchical codebook search is used to improve the efficiency of the algorithm in terms of average number of pilots while also improving the MSE, which are both used to evaluate the performance of the algorithm. The performance of the algorithm is compared with the corresponding performance of the fast channel tracking (FCT) algorithm.

\subsubsection*{Notations} 
Unless otherwise stated, lower case and upper case bold letters denote a vector and a matrix, respectively, while $\mathbf{A}^H$ stands for the conjugate transpose, and $tr(\mathbf{A})$ represents the trace of matrix $\mathbf{A}$. Finally, $\mathbf{I}_{K}$ is the $K \times K$ identity~matrix.

\section{System and signal model}\label{sec:SSM}

\iffalse
\begin{figure}
\centering
\captionsetup{justification=centering}
\includegraphics[width=0.9\linewidth,trim=0 0 0 0,clip=false]{images/SystemModel_Tracking.eps}
\caption{System model.}
\label{fig:SM}
\end{figure}
\fi

\begin{figure}
\centering
\captionsetup{justification=centering}
\includegraphics[width=0.9\linewidth,trim=0 0 0 0,clip=false]{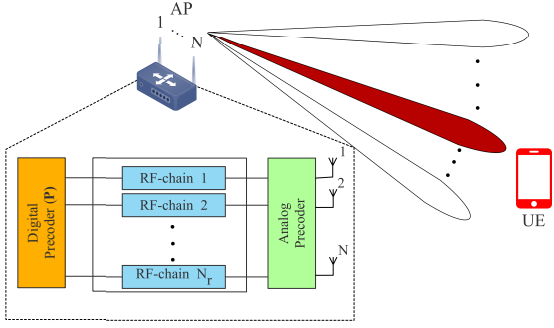}
\caption{System model.}
\label{fig:SM}
\end{figure}

As illustrated in Fig.~\ref{fig:SM}, a typical THz massive multiple-input multiple-output (MIMO) system is considered. The AP is equipped with a uniform linear array (ULA) consisting of $N$ elements that are driven by $N_r$ radio frequency (RF) chains, with $N_r\leq N$, in order to serve $K$ single-antenna UEs. In the downlink, the baseband equivalent received signal vector for all the $K$ UEs can be obtained as~\cite{Gao2017}
\begin{align}
\mathbf{y} = \mathbf{H}^{H}\mathbf{W}^{H}\mathbf{P} \mathbf{s} + \mathbf{z},
\end{align}    
where $\mathbf{s}$, $\mathbf{P}$ and $\mathbf{W}$ respectively stand for the transmitted signal vector for all the $K$ UEs with normalized power, i.e. $\mathbb{E}\left[\mathbf{s}\mathbf{s}^{H}\right]=\mathbf{I}_K$, the digital precoding matrix and the multi-resolution codebook matrix, while $\mathbf{H}$ and $\mathbf{z}\sim\mathcal{CN}(0, \sigma^2 \mathbf{I}_K)$ denote the MIMO channel matrix and the additive Gaussian noise (AWGN) vector with variance $\sigma^2$, respectively. Of note, the precoding matrix satisfies the total transmit power constraint, i.e. $\mathrm{tr}\left(\mathbf{P}\mathbf{P}^{H}\right)\leq P$, with $P$ being the total transmit power. 

\begin{figure}
\centering
%\captionsetup{justification=centering}
\begin{subfigure}[t]{1\linewidth}
\includegraphics[width=1\linewidth,trim=0 0 0 0,clip=false]{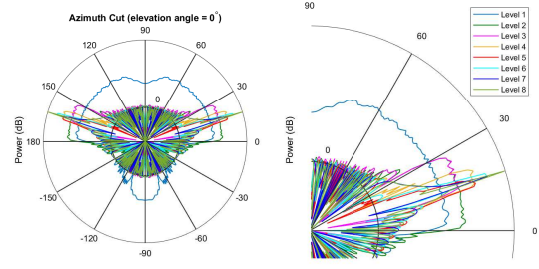}
\vspace{-1cm}
\caption{Patterns of matching codewords between the different levels}
\end{subfigure}
\begin{subfigure}[t]{1\linewidth}
\includegraphics[width=1\linewidth,trim=0 0 0 0,clip=false]{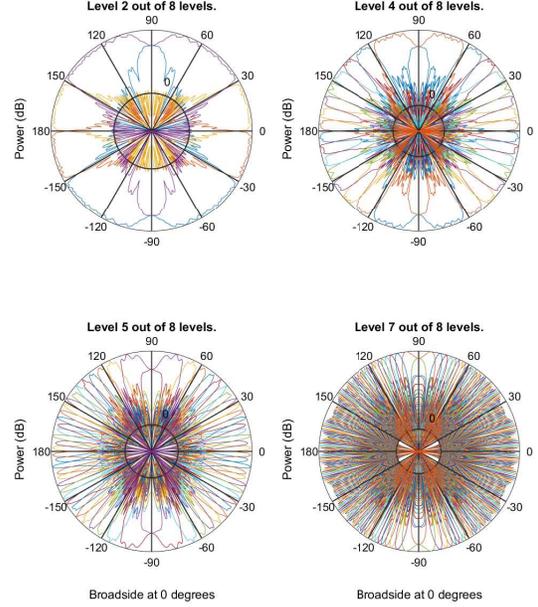}
\vspace{-1cm}
\caption{Patterns of all the codewords within a particular level}
\end{subfigure}
\caption{Indicative example of the codebook design.}
\label{fig:Codebook}
\end{figure}

Without loss of generality, the multi-resolution codebook designed in~\cite[Sec.III.C.3]{A:Hierarchical_codebook_design_for_beamforming_training_in_mmW_communication} has been considered. Note that this codebook is suitable for an analog beamforming architecture and results in  a number of levels equal to $U=\log_2\left(N\right)$, with $2^u$ codewords at each level, where $u=1,2, \cdots, U$. According to~\cite[Cor. 1]{A:Hierarchical_codebook_design_for_beamforming_training_in_mmW_communication}, after the evaluation of the first codeword in each level, the rest of the codewords can be calculated through rotation. An indicative example of the codebook's levels and the corresponding antenna patterns for $N=256$ and center frequency set to $275\text{ }\mathrm{GHz}$ is illustrated in Fig.~\ref{fig:Codebook}.

The MIMO channel matrix can be expressed as $\mathbf{H}=\left[\mathbf{h}_1, \mathbf{h}_2, \cdots, \mathbf{h}_K\right]$, where $\mathbf{h}_k$ is the channel vector corresponding to the $k-$th UE, with $k=1, 2, \cdots, K$. Adopting the widely-used Saleh-Valenzuela model for the THz channel\footnote{Notice that the SV model has been extensively used to model indoor environments in wireless THz systems~(see e.g. \cite{Gao2017,Sayeed2013} and references therein).}, ~$\mathbf{h}_k$ can be obtained~as   
\begin{align}
\mathbf{h}_{k} = \beta_{k}^{(0)} \mathbf{a}(\psi_k^{(0)})+ \sum_{i=1}^{L} \beta_k^{(i)} \mathbf{a}(\psi_k^{(i)}),
\label{Eq:ch_vec}
\end{align}
where $\beta_{k}^{(0)}$ and $\beta_k^{(i)}$ denote the complex gains of the LoS and non-LoS components, respectively, $\mathbf{a}\left(\psi\right)$ stands for the steering vector in the spatial direction $\psi$, while $\psi_k^{(0)}$ and $\psi_k^{(i)}$ respectively represent the spatial directions of the LoS and non-LoS components. In the THz band, scattering induces more than $20\text{ }\mathrm{dB}$ attenuation in the non-LoS components \cite{Papasotiriou2018,A:Analytical_Performance_Assessment_of_THz_Wireless_Systems, C:Analytical_performance_evaluation_of_THz_Wireless_Fiber_Externders,
	our_spawc_paper_2018, Boulogeorgos2019,Boulogeorgos2018a}; hence, only the LoS component can be used reliably. Therefore,~\eqref{Eq:ch_vec} can be simplified~as $\mathbf{h} \approx \beta \mathbf{a}(\psi)$.

\iffalse
\begin{align}
\mathbf{h} \approx \beta \mathbf{a}(\psi).
\label{Eq:ch_vec2}
\end{align}
\fi

The array steering vector can be expressed~as
\begin{align}
\mathbf{a}(\psi)= \dfrac{1}{\sqrt{N}} [e^{-j2\pi \psi m}]_{m\in \mathcal{I}(N)},
\label{Eq:ch_vec2}
\end{align}
where $\mathcal{I}(N)=l-(N-1)/2$, with $l=0,1,...,N-1$ stands for a symmetric set of indices centered around zero. Moreover, the spatial direction is connected with the wavelength $\lambda$ and the inter-element antenna spacing $d$ through
$\psi \triangleq \dfrac{d}{\lambda} \sin\left(\theta\right)$, where $\theta$ is the actual direction of the UE. Finally, without loss of generality, it is assumed that $d= \lambda/2$.

\iffalse
\begin{figure}
\centering
\captionsetup{justification=centering}
\includegraphics[width=0.5\textwidth]{Figures/Results/Random_motion/Possible_starting_locations.eps}
\caption{Possible starting locations in a $5$x$5 m^2$ room.}
\label{fig:Psl}
\end{figure}
\fi

\begin{figure}
\centering
\captionsetup{justification=centering}
\includegraphics[width=0.39\textwidth]{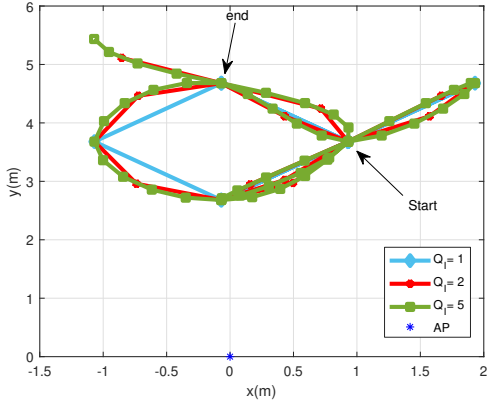}
\caption{Random motion with different $Q_I$.}
\label{fig:Rm}
\end{figure}

%Finally, note that the motions that are used in this paper, are produced as a random walk process \cite{RandWalk2D}. The starting locations are generated uniformly on a disk with center the center of the room. Then, the motions of the UE are generated by the random walk process, and afterwards linear interpolation with $Q_I$ intermediate points is utilized to create the possible intermediate positions of the UE. In other words, $Q_I$ represents the number of direction estimations that the AP performs. The increase of $Q_I$ is expected to reduce the average number of pilots and MSE. Fig. \ref{fig:Rm} shows an example of a random motion with different $Q_I$ values of the linear interpolation.
%The case with $Q_I= 1$ is the motion generated by the random walk process.

Note that the UE motions is modeled as a random walk process\cite{RandWalk2D}. The starting locations of the UE are generated uniformly on a disk centered at the room center. The motion of the UE is generated by the random walk process, and is resampled with a factor $Q_I>1$. The number of timeslots of the resulting motion is $Q_I$ times greater than the original's ($Q_I= 1$). The increase of $Q_I$ is expected to reduce the average number of pilots and MSE. Fig. \ref{fig:Rm} shows an example of a random motion with different factors $Q_I$. 

\section{Hierarchical beam-tracking approach}

For the sake of simplicity, a single UE is considered, i.e. K=1. The proposed approach consists of three phases, namely initialization, location prediction, and direction refinement.

\subsection{Phase 1: Initialization}
The objective of this phase is to accurately determine three consecutive UE locations, in order for Phase 2 to begin and reduce the pilot overhead to affordable levels. In this sense, during the first three timeslots, a combination of high and low level codewords is used. In each one of these timeslots, the following procedure is~employed:  
\begin{enumerate}
\item \underline{Level 1 codewords exhaustive search:} The AP performs an exhaustive search by switching between the codewords of the first level, which generate the lowest available resolution antenna pattern, in order for the AP to determine the sector in which the UE is located, i.e. the level 1 codeword that should be used to serve the~UE.
\item \underline{Higher level codeword refinement:} Afterwards, the AP refines the search by selecting the next higher level of the codebook and switching between those codewords of this level that generate beam patterns within the direction specified in the previous step. Note that these codewords generate narrower beams in comparison with those of the previous level. This step is repeated until the algorithm reaches the highest level of the codebook. If the algorithm fails to find the direction of the UE for a codebook level, it switches to one level lower and continues the procedure.\\
%In this step, the codeword that results in the highest signal-to-noise ratio (SNR), regardless of the codebook level, can be saved for use in possible transmissions. This can be useful in cases where a lower level codeword of the lower levels of the codebook can generate a beam that points closer to the UE's actual direction than the higher level one and the AP-UE distance is short enough for the lower antenna gain to be adequate for reliable reception.
\item \underline{Distance and location estimation:} To estimate the transmission distance a two-way time of arrival approach \cite{Dargie2010} has been employed. The AP-UE distance can be obtained as
\begin{align}
% p = \frac{\left(t_r^u-t_r^a\right)-\left(t_s^u-t_s^a\right)}{2} c,
\tilde{r}= r+ e,
\end{align}       
%where $c$ stands for the speed of light.
where $r$ is the actual AP-UE distance  and $e$ is a normally distributed random variable with zero mean and standard deviation $\sigma_e$ \cite{Kim2010}. By defining a two-dimensional Cartesian coordinate system, in which the AP is placed at $\left(0,0\right)$, as illustrated in Fig. 3, the location of the UE can be estimated~as
\begin{align}
\mathbf{p}(x, y) = \left(\tilde{r} \cos\tilde{\theta}, \tilde{r} \sin\tilde{\theta}\right),
\end{align} 
where $\tilde{\theta}$ is the main lobe direction of the highest level~codeword and represents the estimated direction of the UE, while $x$ and $y$ stand for the coordinates.
\end{enumerate}

\subsection{Phase 2: Location prediction}
After obtaining the UE's three consecutive locations from Phase 1 and assuming that the UE follows a linear motion, the AP can predict the UE's upcoming location as \cite{Stratidakis2019}
\begin{gather}
	\begin{split}
	&\mathbf{p}(t+1)= \mathbf{p}(t)+ \\
	&\dfrac{\ [\mathbf{p}(t)-\mathbf{p}(t-1)]+[\mathbf{p}(t-1)- \mathbf{p}(t-2)]}{2},
	\end{split}
	\label{Eq:pos_pred}
\end{gather}
where $t$ refers to the current timeslot. The accuracy of the prediction helps reduce the required pilot overhead as fewer codewords need to be tested by the AP. Therefore, the required pilot overhead can be reduced significantly.

\subsection{Phase 3: Direction refinement}
In this Phase, the AP refines the prediction of Phase 2. The AP switches between the codewords of level U-g, where $1 \leq g\leq U-3$, that generate beams toward and around the predicted direction, in order to find the UE's direction. Next, it uses the next level to find a more accurate direction, by generating beams towards the direction pointed by the previous level. This is repeated until the last codebook level. The use of multiple codebook levels in this step is expected to increase the robustness of the algorithm, in cases of non-linear motions and non-constant speed, while keeping the pilot overhead low. Finally, the AP repeats step 3 of Phase 1, Phase 2 and Phase 3.

%The AP selects the U-g codebook level, where $1 \leq g\leq U-3$, and switches between its codewords that generate beams toward and around the predicted direction in order to find the UE's direction. 

\section{Simulation Results \& Discussion}
In this section, the effectiveness of the proposed approach is evaluated in terms of average number of pilots and MSE= $\dfrac{1}{n}\sum_{i=1}^{n}(\tilde{\theta}-\theta)^2$. The results are derived by means of Monte Carlo simulations in 10,000 different motions generated by the new motion generation. The following insightful scenario is considered. The room dimension is $5$x$5$ $m^2$, the step of the random motion is $1\text{ }m$ and the number of timeslots is 10 times the value of $Q_I$. The ULA consists of 256 elements; thus, $8$ codebook levels are considered, while the carrier frequency and the signal-to-noise ratio (SNR) are equal to 275 GHz and 10 dB, respectively. The FCT algorithm that was presented in \cite{Gao2017}, is used as a benchmark. Notice that FCT essentially employs the last codebook level and a prediction based on previous directions. Moreover, it uses $128$ pilots in the first $3$ timeslots that compose the initialization phase and $16$ pilots thereafter. Unless otherwise stated, the number of pilots per codebook level of the proposed approach is $2$, in the first $2$ levels, and $4$ thereafter. Therefore, the total number of pilots in Phase 1 is $28$. Phase 3 utilizes the last 4 codebook levels, which result in $16$ pilots. Finally, the algorithms restart in the next timeslot if $|\tilde{\theta}(t)-\theta(t)|> \theta_{3dB}$, where $\theta_{3dB}$ is the half power beamwidth. Notice that if this condition holds, the communication between the AP and the UE is interrupted. 

%If the algorithm fails to find the UE's direction, as the received signal-to-noise ratio (SNR) at the estimated direction is $0$,
%($|\tilde{\theta}-\theta|< \theta_{3dB}$), 
%it restarts in the next timeslot in order to find the UE.

\begin{figure}
\centering
        \includegraphics[width=0.39\textwidth]{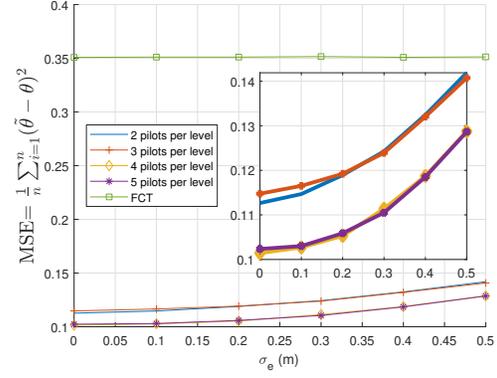}
        \caption{MSE vs $\sigma_e$.}
        \label{fig:MSEvss}
\end{figure}

In Fig. \ref{fig:MSEvss}, the MSE is illustrated as a function of $\sigma_e$, for different number of pilots per codebook level and $Q_I= 10$. The number of pilots per codebook level affects the number of directions searched; hence, the area covered by each level of the codebook during tracking. As a consequence, a low number of pilots covers a small area and limits the effectiveness of tracking in terms of MSE. On the other hand, as the number of pilots increases, the overall tracking overhead also increases. Moreover, from this figure, it is evident that independently of $\sigma_e$, approximately the same error performance can be achieved with $4$ or $5$ pilots per level. Additionally, as the number of pilots per level increases from $3$ to $4$, MSE decreases for about 0.0133. Finally, the MSE of FCT is $3$ times higher than the proposed approach with $\sigma_e= 0$ $m$ and $2.5$ times higher with $\sigma_e= 0.5$ $m$.

\begin{figure}
\centering
        \includegraphics[width=0.39\textwidth]{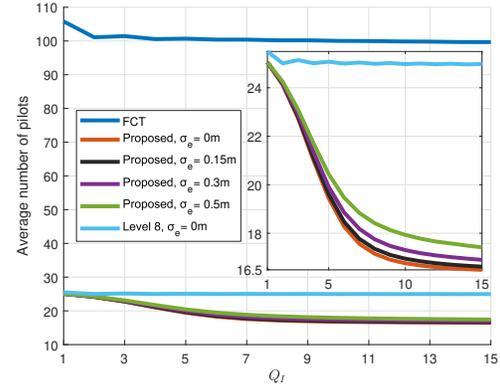}
        \caption{Average number of pilots vs $Q_I$.}
        \label{fig:Anopvsfoe}
\end{figure}

Fig. \ref{fig:Anopvsfoe} depicts the average number of pilots versus $Q_I$, for different $\sigma_e$ values. The graph denoted as ``Level 8" corresponds to the special case in which only the last codebook level is employed in Phase 3 of the proposed approach with the same total number of pilots. From this figure, we observe that, for a given $\sigma_e$, the average number of pilots, when the proposed approach is employed, decreases as $Q_I$ increases. On the other hand, the average number of pilots of the ``Level 8" approach and the FCT is almost constant and independent of $Q_I$. The high average number of pilots of the FCT and ``Level 8" approach is related to the estimation error as well as the higher number of pilots that are employed during the initialization phase of FCT. Notice, that according to this result, the average number of pilots for the FCT is between 99.7 and 105.8, while for the ``Level 8" approach it is between 25 and 25.5. On the contrary, the average number of pilots for the proposed approach is between 16.5 and 25 with $\sigma_e= 0$ $m$, while it is between 17.4 and 25 with $\sigma_e= 0.5$ $m$. Thus, the use of a hierarchical codebook, makes tracking more robust to direction and speed variations.

\section{Conclusion}
In this paper, a novel hierarchical beam-tracking algorithm for indoor THz communication systems that requires low pilot overhead and provides high tracking efficiency was presented. The proposed approach requires $78.13\%$ less pilot overhead in the initialization phase than the FCT, while the overall overhead reduction that can be achieved by the proposed approach in comparison with the FCT may exceed  $76\%$. Finally, it achieves lower error performance compared to FCT.

\section*{Acknowledgment}
This work has received funding from the European Commission's Horizon 2020 research and innovation programme TERRANOVA under grant agreement No. $761794$ (TERRANOVA) and No. $871464$ (ARIADNE).

\balance
\bibliographystyle{IEEEtran}
\bibliography{IEEEabrv,References}

\end{document}